\documentclass[aps,prd,preprintnumbers,twocolumn,groupedaddress,nofootinbib]{revtex4}

\pdfoutput=1

\usepackage{graphicx}
\usepackage{latexsym}
\usepackage{amsfonts}
\usepackage{amssymb}
\usepackage{amsmath}
\usepackage{slashed}
\usepackage{array}
\usepackage{feynmp}
\usepackage{url}

\begin{document}
\title{Precision Probes of a Leptophobic $Z'$ Boson}

\author{Matthew R.~Buckley$^{1}$ and Michael J. Ramsey-Musolf$^{2,3}$}
\affiliation{$^1$Center for Particle Astrophysics, Fermi National Accelerator Laboratory, Batavia, IL 60510, USA}
\affiliation{$^2$Department of Physics, University of Wisconsin, Madison, WI 53706, USA}
\affiliation{$^3$Kellogg Radiation Laboratory, California Institute of Technology, Pasadena, CA 91125, USA}

\preprint{FERMILAB-PUB-12-049-A}

\preprint{NPAC-12-03}
\date{\today}

\begin{abstract}
Extensions of the Standard Model that contain leptophobic $Z'$ gauge bosons are theoretically interesting but difficult to probe directly in high-energy hadron colliders. However, precision measurements of Standard Model neutral current processes can provide powerful indirect tests. We demonstrate that parity-violating deep inelastic scattering of polarized electrons off of deuterium offer a unique probe leptophobic $Z'$ bosons with axial quark couplings and masses above 100~GeV. In addition to covering a wide range of previously uncharted parameter space, planned measurements of the deep inelastic parity-violating $eD$ asymmetry would be capable of testing leptophobic $Z'$ scenarios proposed to explain the CDF $W$ plus di-jet anomaly.
\end{abstract}

\maketitle

The addition of a new abelian gauge group is one of the simplest extensions to the Standard Model (SM) that can be considered. In principle, a completely generic $U(1)'$ and its associated gauge boson, the $Z'$, could have arbitrary generation-dependent couplings to the known particles, with the resulting triangle anomalies cancelled by the addition of new heavy chiral fermions. The resulting embarrassment of (theoretical) riches arising from this freedom of choice calls for the addition of a guiding symmetry principle as to impose some amount of order.  Widely considered gauge groups (see, for example the reviews Refs.~\cite{Nakamura:2010zzi,Langacker:2008yv} and references therein) include gauged $B-L$ (the unique choice that is anomaly free with the Standard Model fermion content), $B-xL$ with $x$ a free parameter, Grand Unified Theory (GUT)-derived models, and leptophilic $Z'$ bosons. The latter have held particular interest recently in the context of explaining the PAMELA \cite{Adriani:2008zr} and Fermi \cite{Abdo:2009zk} anomalies in terms of dark matter \cite{ArkaniHamed:2008qn,Essig:2009nc}. 

The majority of the models that have been studied contain sizable couplings to leptons -- an important feature as the dominant experimental constraints come from processes involving leptons (for a recent global analysis, see Ref.~\cite{Erler:2011iw}). For example, a sequential $Z'$, whose  couplings to SM fermions are proportional to those of the $Z$, is ruled out  for $M_{Z'}$ below $\sim 1$~TeV. Similar constraints hold for other scenarios with leptonic couplings \cite{Nakamura:2010zzi}. Intriguingly, $Z'$ bosons that couple exclusively (or at least predominantly) to quarks are not as strongly limited by collider experiments, due to the large QCD backgrounds. The most obvious channel for a leptophobic $Z'$ search at hadronic machines, $p\bar{p}/pp \to Z' \to j j$, is stymied at low $Z'$ mass by the prohibitive dijet background rate. Currently, the tightest bound in this channel for a $Z'$ below $\sim 300$~GeV with electroweak-scale couplings comes from the UA2 experiment \cite{Alitti:1993pn} (see also Refs.~\cite{Buckley:2011vc,Yu:2011cw}).

In the last year, the CDF collaboration reported an excess of events in the $W^\pm+jj$ channel, seen as a Gaussian peak in the $m_{jj}$ distribution at $147\pm4$~GeV \cite{Aaltonen:2011mk}. This anomaly, initially reported at $3.2\sigma$ in 4.3~fb$^{-1}$, growing to $4.2\sigma$ in 7.3~fb$^{-1}$ \cite{CDF2}, can be interpreted as a new $Z'$ coupling to quarks with a mass of $\sim 150$~GeV and a charge times gauge coupling of ${\cal O}(0.2-0.5)$ \cite{Buckley:2011vc,Yu:2011cw,Cheung:2011zt}. Particular theoretical realizations of such leptophobic $Z'$ models have since been considered; for example, separately gauged $B$ and $L$ \cite{Buckley:2011vs}, or a $E_6$ GUT with hypercharge-$U(1)_\eta$ mixing \cite{Buckley:2011mm}. A D\O\ search does not see a similar excess \cite{Abazov:2011af}, and disagreement between the two experiments remains. The situation is unlikely to be fully resolved until ATLAS and CMS weigh in with $5-10$~fb$^{-1}$ of data \cite{Eichten:2011xd,Buckley:2011hi}.

Regardless of the final resolution of this particular anomaly, it is clear that leptophobic gauge groups are both theoretically interesting and not well constrained by existing searches. In this paper, we propose a new precision probe of leptophobic $Z'$ bosons using parity-violating deep inelastic scattering (PV-DIS) of electrons off of deuterium. Historically, PV-DIS played a key role in singling-out the Glashow-Weinberg-Salaam prediction for the neutral weak interaction from among alternative possibilities. From a theoretical perspective, it has often been considered as a potentially powerful indirect probe of possible physics beyond the Standard Model (see, {\em e.g.}, \cite{RamseyMusolf:2006vr,Erler:2004cx,Robinett:1981yz} and references therein).
In the present era, a measurement of the PV asymmetry has recently been completed with the GeV beam at Jefferson Lab (JLab) \cite{Subedi:2011zz}, while more precise measurements are planned for the JLab 12 GeV program \cite{SOLID}, and discussed as a possibility for a future Electron Ion Collider (EIC) \cite{Boer:2011fh}. 

As we will show, the future PV-DIS asymmetry measurements would be sensitive to axial couplings to quarks of a $Z'$ with mass and couplings required to explain the CDF anomaly. Furthermore, these measurements would be competitive with the current leading experimental bounds. In what follows, we use the leptophobic $E_6$ model of Ref.~\cite{Buckley:2011mm} as a benchmark scenario, but provide a more general framework for assessing the leptophobic $Z'$ scenario.

The effect of new physics on parity violation in deep inelastic $eD$ scattering is parameterized by four couplings in the effective Lagrangian:
\begin{equation}
{\cal L}_{\rm PV} = \frac{G_F}{\sqrt{2}} \sum_q \left[ C_{1q} (\bar{e} \gamma^\mu \gamma_5 e)( \bar{q}\gamma_\mu q )+ C_{2q} (\bar{e} \gamma^\mu e)( \bar{q}\gamma_\mu \gamma_5 q ) \right]. \label{eq:efflag}
\end{equation} 
Here, the sum is over the valence quarks ($q= u,d$). In the SM, these couplings are (see, {\em e.g.}, Ref.~ \cite{RamseyMusolf:2006vr})
\begin{eqnarray}
\label{C12rad}
C_{1q} & = & 2 {\hat\rho}_{NC} I_3^e \left( I_3^q-2 Q_q {\hat\kappa}\sin^2{\hat\theta}_W\right) -\frac{1}{2}{\hat\lambda}_1^q \label{eq:C1} \\ 
C_{2q} & = & 2 {\hat\rho}_{NC} I_3^q \left( I_3^e-2 Q_e {\hat\kappa}\sin^2{\hat\theta}_W\right) -\frac{1}{2}{\hat\lambda}_2^q\ \ \ , \label{eq:C2}
\end{eqnarray}
where $I_3^f$ is the third component of weak isospin for fermion $f$, $Q_f$ is the electromagnetic charge, and ${\hat\theta}_W$ angle is the weak mixing in the $\overline{\mathrm{MS}}$ scheme. The quantities $ {\hat\rho}_{NC}$, ${\hat\kappa}$, and ${\hat\lambda}_j^q$ encode the effects of electroweak radiative corrections and at tree-level take on the values 1, 1, and 0, respectively. Theoretically, the $C_{1q}$ and $C_{2q}$ are predicted to better than one percent precision. Experimentally, the nuclear weak charge  $Q_W=-2(Z C_{1u}+ N C_{1d})$ has been determined at the $\sim 0.5\%$ level by measurement of PV transitions in cesium \cite{Wood:1997zq,Porsev:2009pr}, while the proton weak charge $Q_W^p=-2(2 C_{1u}+ C_{1d})$ will be determined to $~ 4\%$ precision with PV elastic $ep$ scattering at JLab by the Q-Weak Collaboration \cite{QWeak}. Note that at tree level $Q_W^p=1-4\sin^2{\hat\theta}_W\sim 0.1$, so that a 4\% determination of this quantity is roughly comparable to a 0.5\% determination of the cesium weak charge. For a summary of present and prospective constraints on the $C_{1q}$ see Ref.~\cite{Young:2007zs}. 

In contrast, the present experimental bounds on the $C_{2q}$ are considerably weaker, a situation that would be remedied by the PV-DIS studies. Experimentally, the projected precision of the SOLID experiment would yield a determination of $2C_{2u} - C_{2d}$ with an uncertainty $ \pm 0.0083$ \cite{SOLID}. An EIC measurement could lead to a factor of two-to-three smaller uncertainty, provided an ultra-high luminosity version is ultimately constructed, with an integrated luminosity of  0.5 to 1 attobarn$^{-1}$ \cite{KK}.

The PV $eD$ asymmetry is sensitive to both the $C_{1q}$ and $C_{2q}$:
\begin{eqnarray}
\label{CG-modify}
A_{PV}^{eD} &=& - \frac{G_\mu Q^2}{2\sqrt{2}\pi \alpha}\frac{9}{10}\Big [\tilde{a}_1  + \tilde{a}_2 \frac{1-(1-y)^2}{1+(1-y)^2}\Big ],
\end{eqnarray}
where $G_\mu$ is the Fermi constant as determined from the muon lifetime, the parameter $-Q^2=q^2=q_0^2-|{\vec q}|^2$ is the square of the four momentum transfer, and the $\tilde{a}_{1,2}$ are given by
\begin{eqnarray}
\label{R-effects}
\tilde{a}_1 &=& -\frac{2}{3}\left(2C_{1u}-C_{1d}\right) \big [1 + R_1  \big ], \nonumber \\
\tilde{a}_2 &=& -\frac{2}{3}\left(2C_{2u}-C_{2d}\right) \big [1  + R_2 \big ]. \nonumber
\end{eqnarray}
Here the $R_k$ denote various hadronic corrections, including those associated with higher twist contributions to the deep inelastic structure functions and charge symmetry violation (CSV) in the parton distribution functions (for recent discussions, see Refs.~\cite{Mantry:2010ki,Hobbs:2008mm}). Through an appropriate program of measurements at different kinematics ($Q^2$ and Bjorken-$x$), it is in principle possible to disentangle these hadronic contributions from the $Q^2$- and $x$-dependent terms. 

In general, new physics could become apparent in both $C_{1q}$ and $C_{2q}$. Given the sensitivity of the cesium atomic PV and Q-Weak experiments to the $C_{1q}$, it is relevant to ask what complementary information a determination of the $C_{2q}$ coefficients from $A_{PV}^{eD}$ might provide. In this context, the leptophobic $Z'$ scenario is particularly interesting, as it will not affect the $C_{1q}$ at an appreciable level but could lead to a sizeable shift in the $C_{2q}$ as we show below. 

\begin{figure}[t]
\includegraphics[width=0.7\columnwidth]{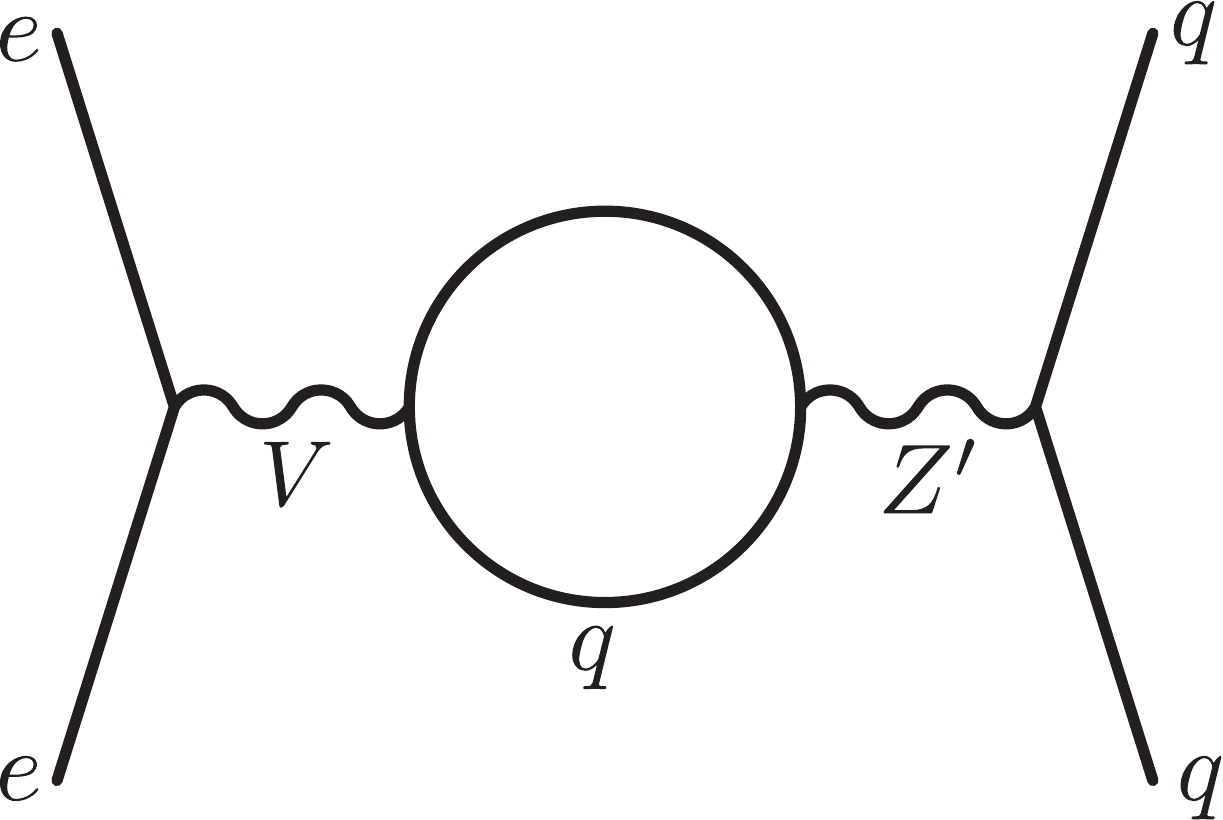}
\caption{Loop diagram leading to corrections to the coefficients $C_{1q}$ and $C_{2q}$ in Eq.~\eqref{eq:efflag} due to a new $Z'$ gauge boson coupling exclusively to quarks. In general, the vector boson $V$ can be either $\gamma$ or $Z$. Requiring photon coupling to electrons, axial couplings of the $Z'$ will result in corrections to $C_{2q}$. \label{fig:loop}}
\end{figure}

Since (by assumption) the $Z'$ does not couple to the electrons, its dominant contribution to the $({\bar e}\gamma^\mu e)\,({\bar q}\gamma_\mu\gamma_5 q)$ operator arises at one-loop level through $\gamma Z'$ mixing tensor as shown in Fig.~\ref{fig:loop}. The leptophobic $Z'$ couples only to quarks in the loop, in contrast to analogous $\gamma Z$ mixing in the SM that also includes lepton loops. The corresponding effect does not enter the $({\bar e}\gamma^\mu \gamma_5 e)\, ({\bar q}\gamma_\mu q)$ operator proportional to $C_{1q}$ as the photon has no tree-level axial coupling to the lepton and since the $eeZ'$ vertex vanishes. In principle, the analogous process involving  $Z-Z'$ mixing would lead to shifts in both $C_{1q}$ and $C_{2q}$. However, the mixing angle $\alpha_{ZZ'}$ is constrained to be $\lesssim 10^{-3}$ \cite{Erler:2009ut}, rendering the effect too small to be observable in the next generation of experiments.\footnote{The specific mechanism for ensuring sufficiently small $Z-Z'$ mixing requires a detailed discussion of the scalar sector of the $U(1)^\prime$ extension, a topic that goes beyond the scope of the present work. See {\em e.g.} Refs.~\cite{Hewett:1985ss,Hewett:1986bk,Cvetic:1997ky,Amini:2002jp} and references therein for treatments within the context of supersymmetric $U(1)^\prime$ models.}

In what follows, we illustrate the prospective sensitivity of the PV-DIS asymmetry to $\gamma Z'$ exchange. We observe that the expected shift $\Delta C_{2q}$ is enhanced relative to the na\"ive expectation of $(\alpha/\pi) (M_Z/M_{Z'})^2$ by two effects: the sum over quark colors and the presence of large logarithms that arise at the relatively low-$Q^2$ of the PV-DIS experiments. In addition, the SM predictions for the $C_{2q}$ are suppressed, as the tree-level values are proportional to $1-4\sin^2{\hat\theta}_W$, leading to an additional transparency to a $\gamma Z'$ mixing contribution that does not carry this suppression factor. 

We first review the computation of the tree-level contribution to coefficient $C_{2q}$ that arises from $eq$ scattering via a SM $Z$-boson. We define the axial and vector couplings to the $Z$ and $Z'$ gauge bosons via the Lagrangian
\begin{eqnarray}
{\cal L} & = & \sum_f \frac{g}{\cos\theta_W}\bar{f}\gamma^\mu \left[Q_{V,f}+Q_{A,f}\gamma_5 \right] f Z_\mu \nonumber \\
 & & + g' \bar{f} \gamma^\mu \left[ Q'_{V,f} +Q'_{A,f} \gamma_5 \right] f Z'_\mu, \\
Q_{V,f} & = & \frac{1}{2}\left(I_{3,f}-2Q_f\sin^2\theta_W \right), \\
Q_{A,f} & = & -\frac{1}{2} I_{3,f}\ \ \ ,
\end{eqnarray}
where we have dropped the hat notation from Eqs.~\eqref{eq:C1} and \eqref{eq:C2} for simplicity. Here, the coupling $g$ is the $SU(2)_L$ gauge coupling, while the new gauge coupling $g'$ and charges $Q'_{A,f}$ and $Q'_{V,f}$ are model dependent. 

In terms of the vector and axial charges to electrons ($Q_{V,e}$ and $Q_{A,e}$) and quarks ($Q_{V,q}$ and $Q_{A,q}$), the scattering matrix element is
\begin{eqnarray}
i{\cal M}_{\rm tree} & = & \frac{ig^2}{\cos^2\theta_W} \frac{1}{q^2-M_Z^2}  \times \label{eq:tree1} \\
 & &[ \bar{e} \gamma^\mu(Q_{V,e}+Q_{A,e}\gamma_5) e ][\bar{q} \gamma_\mu (Q_{V,q} + Q_{A,q}\gamma_5)q] .\nonumber
\end{eqnarray}
Taking the $q^2 \to 0$ limit, comparing to the effective Lagrangian in Eq.~\eqref{eq:efflag}, and using  $G_\mu/\sqrt{2} \equiv g^2/8M_Z^2 \cos^2\theta_W$, leads to the tree-level $C_{2q}$:
\begin{eqnarray}
C_{2q} & = & -8 Q_{V,e} Q_{A,q} \nonumber \\
 & = & 2I_{3,q}\left(I_{3,e}-2Q_e\sin^2\theta_W \right), \label{eq:treec2q} \\
C_{2u} & = & -\frac{1}{2}\left(1-4\sin^2\theta_W\right) = -0.0372, \label{eq:treec2u}\\
C_{2d} & = &  +\frac{1}{2}\left(1-4\sin^2\theta_W\right) =+0.0372. \label{eq:treec2d}
\end{eqnarray}
Including the electroweak radiative corrections $\overline{\mathrm{MS}}$ scheme indicated in Eq.~(\ref{C12rad}),  one obtains $C_{2u} = -0.0357$, $C_{2d}=0.0268$~\cite{Nakamura:2010zzi}, yielding $2C_{2u}-C_{2d} = -0.0981$. Thus, the projected sensitivity on $2C_{2u}-C_{2d}$ of the SOLID experiment is approximately $8.5\%$ of the SM value.

A substantial  contribution to the SM corrections arises from $\gamma Z$ mixing that enters the quantity ${\hat\kappa}$ in Eq.~(\ref{C12rad}). This quantity depends on both $Q^2$ and the t'Hooft (renormalization) scale $\mu$, while the product ${\hat\kappa}(Q^2, \mu)\sin^2{\hat\theta}_W(\mu)$ is $\mu$-independent. Choosing $\mu= M_Z$, as is appropriate when comparing to $Z$-pole precision observables ($Q^2=-M_Z^2$), we encounter large logarithms in the theoretical predictions for the low-$Q^2$ asymmetries of interest here. In this case, renormalization group (RG) improved predictions can be obtained by choosing $\mu\sim \sqrt{Q^2}$ and exploiting the RG evolution of $\sin^2{\hat\theta}_W(\mu)$ as discussed in Ref.~\cite{Erler:2004in}. Doing so resums the large logarithms by moving them from ${\hat\kappa}(Q^2, \mu)$ into $\sin^2{\hat\theta}_W(\mu)$.

Next, we consider the $\gamma Z'$ contribution. For purposes of illustrating the magnitude of this effect, we will defer a full RG-improved analysis to future work, concentrating instead on  the $\gamma Z'$ contribution to ${\hat\kappa}(Q^2,M_Z)$ given its conceptual simplicity.  Following the approach of Ref.~\cite{Marciano:1983ss},   we define for general gauge bosons $V$ and $V'$
\begin{eqnarray}
\Pi_{VV'}^{\mu\nu}(q^2) & = & \left. i\int d^4x e^{-iq\cdot x}\langle 0 | {\hat T} J_V^\mu(x)J_{V'}^\nu(0) | 0\rangle\right|_T \label{eq:Pidef} \\
\Pi_{VV'}^{\mu\nu}(q^2) & = &  \left(q^\mu q^\nu  - q^2 g^{\mu\nu} \right)\Pi_{VV'}(q^2), \label{eq:Pidef2}
\end{eqnarray}
where the ${\hat T}$ is the time-ordering operator, $J_V^\mu$ ($J_{V'}$) is the current that couples to vector boson $V$ ($V'$),  and the subscript \lq\lq $T$" denotes the transverse component. With this normalization, the matrix element for $eq$ scattering via the loop diagram shown in Fig.~\ref{fig:loop} is given by
\begin{equation}
i{\cal M} = ie g' \frac{ \Pi_{\gamma Z'}(q^2)}{q^2-M_{Z'}^2} [\bar{e}\gamma^\mu e][\bar{q}\gamma_\mu(Q_{V,q}'+Q_{A,q}'\gamma_5)q]. \label{eq:loop1}
\end{equation}
Again taking the low $q^2$ limit and factoring out $G_\mu/\sqrt{2}$ in order to compare with Eq.~\eqref{eq:efflag}, we find
\begin{eqnarray}
i{\cal M}_{\gamma Z'}^{PV} & = & i\frac{G_\mu}{\sqrt{2}} \left[8 \cos^2\theta_W\sin\theta_W \left( \frac{g'}{g}\right) \left(\frac{M_Z}{M_{Z'}}\right)^2 Q_{A,q}' \right]  \nonumber \\
 & \times& \Pi_{\gamma Z'}(q^2) [ \bar{e} \gamma^\mu e ][\bar{q} \gamma_\mu \gamma_5 q] . \label{eq:PVloop}
\end{eqnarray}

We  now turn to calculating $\Pi_{\gamma Z'}(q^2)$. For heavy quarks ($q=c$, $b$, $t$), the one-loop perturbative calculation yields a reliable result \cite{Marciano:1983ss}:
\begin{eqnarray}
\label{eq:piQ}
[\Pi_{\gamma Z'}(0)]_{c,b,t} &=& -N_c \frac{eg'}{2\pi^2}\sum_q Q_q Q_{V,q}' F(m_q^2,Q^2), \\
\nonumber
F(m_q^2,Q^2)&=&\int_0^1dx\ x(1-x) \ln\left[\frac{m_q^2+x(1-x)Q^2}{M_Z^2}\right],
\end{eqnarray}
where $N_c$ is the number of quark colors.

However, as the light quarks ($u$, $d$, and $s$) have masses at or below the QCD scale, we must take non-perturbative effects into account. Following Ref.~\cite{Marciano:1983ss}, we proceed by splitting the light quark contribution to the $\Pi^{\mu\nu}$ tensor into isovector and isoscalar contributions, leading to :
\begin{eqnarray}
\Pi_{\gamma Z'}(q^2) & = & eg' \Big[ (Q_{V,u}'-Q_{V,d}')\Pi_{I=1}+\label{eq:isospin} \\
 & & \frac{1}{3}(Q_{V,u}'+Q_{V,d}')\Pi_{I=0} + \sum_{q=s,c,b,t} Q_q Q_{V,q}' \Pi_q \Big]\ .  \nonumber
\end{eqnarray}
Note that we have included the top quark in the sum, in contrast to the conventional treatment of $\Pi_{\gamma Z}$ \cite{Marciano:1990dp,Fanchiotti:1992tu}. In the latter instance, one absorbs effects of order $\alpha \ln m_t/M_Z$ in the definition of $\sin^2{\hat\theta}(M_Z)$, a quantity that one extracts from precision $Z$-pole observables. In the $Z^\prime$ case, however, the top contribution to $\Pi_{\gamma Z'}$ induces a non-vanishing $eeZ^\prime$ vector coupling that does not exist at tree-level. Consequently, it is not possible to absorb these loop effects in the definition of renormalized  $Z^\prime$ vector couplings to leptons.

For the three light quarks, data from $e^-e^+$ scattering to hadrons can be used to estimate the $\Pi$ functions at $q^2 = 0$:
\begin{eqnarray}
\Pi_{I=0}(0) = \Pi_{I=1}(0)=  0.178, & & ~ \label{eq:piiso}\\
\Pi_s(0) = 0.292. & & ~ \label{eq:pistrange}
\end{eqnarray}
The $c$ and $b$ quark contributions can be reliably calculated from Eq.~\eqref{eq:piQ}. If we replace $g'$ with $g/\cos\theta_W$ and $Q'_{V,q}$ with the Standard Model $Z$ vector charges $Q_{V,q}$ in Eq.~\eqref{eq:isospin} we reproduce the standard one-loop quark contribution $\Pi_{\gamma Z}$, which contributes to both $C_{2,q}$ and the running of $\sin^2\theta_W$ \cite{Marciano:1983ss,Marciano:1982mm,Czarnecki:1995fw}. 

Combining Eqs.~\eqref{eq:PVloop}-\eqref{eq:pistrange}, we see that, in the $q^2 = 0$ limit, the shift in $C_{2q}$ due to a $Z'$ gauge boson is
\begin{eqnarray}
\Delta C_{2q} & = &32 \pi \alpha \cos^2\theta_W \left( \frac{g'}{g}\right)^2 \left(\frac{M_Z}{M_{Z'}}\right)^2 Q_{A,q}' \times  \label{eq:deltaC} \\
& & \left[ \frac{2}{3}(2Q_{V,u}'-Q_{V,d}')(0.178)-\frac{1}{3}Q_{V,s}'(0.292)+ \right. \nonumber \\
 & & \left. \frac{2}{3} Q_{V,c}' (0.210)-\frac{1}{3}Q_{V,b}' (0.150)-\frac{2}{3}Q'_{V,t}(0.032)\right]. \nonumber
\end{eqnarray}

To investigate the experimental sensitivity to this contribution,  we select as a benchmark model the leptophobic $E_6$ GUT scenario outlined in Ref.~\cite{Barger:1996kr} (and applied to the recent CDF $W^\pm+jj$ excess \cite{Aaltonen:2011mk,CDF2} in Ref.~\cite{Buckley:2011mm}). In this model, the charges of the Standard Model particles are well defined. For the up- and down-type quarks they are:
\begin{eqnarray}
Q_{V,u}' = \frac{1}{6}, & & Q_{A,u}' = -\frac{1}{2} ,\label{eq:E6up} \\
Q_{V,d}' =  -\frac{1}{3}, & & Q_{A,d}' = 0. \label{eq:E6down}
\end{eqnarray}
With this normalization of the charges, in order to explain the overall cross section of the CDF excess, the gauge coupling constant $g'$ must be $\sim 0.6$. The dijet excess is observed at $m_{jj} = 147\pm 4$~GeV, and so for this work we take $M_{Z'} = 150$~GeV. Using these nominal values, in our $E_6$ leptophobic benchmark, we find  at $Q^2=0$
\begin{eqnarray}
(\Delta C_{2u})_{E_6} & = &  -0.0155 \left(\frac{150~\mbox{GeV}}{M_{Z'}}\right)^2\left(\frac{g'}{0.6}\right)^2, \label{eq:C2uE6} \\
(\Delta C_{2d})_{E_6} & = & 0. \label{eq:C2dE6}
\end{eqnarray}
This corresponds to a $\sim 40\%$ correction to the SM value for  $C_{2u}$. The lack of correction to $C_{2d}$ is a model-dependent feature of the leptophobic $E_6$, and is not generically expected of a new $U(1)'$ with axial charges. The future PV-DIS experiments will be carried out at non-zero $Q^2$, so one must evolve the result in Eq.~(\ref{eq:deltaC}) to the appropriate kinematic regime. To that end, we follow Ref.~\cite{WJMSLAC}, and use the perturbative result in Eq.~(\ref{eq:piQ}) with \lq\lq effective" light quark masses:
$m_u=62$ MeV, $m_d=83$ MeV, and $m_s=215$ MeV -- choices that yield a good fit to the dispersive result. For the kinematics of the SOLID experiment, $4\ \mathrm{GeV}^2/c^2 < Q^2 < 10\ \mathrm{GeV}^2/c^2$ this parameterization leads to a reduction in the magnitude of $\Delta C_{2u}$ by $\sim 25 \%$ ($\sim 30 \%$) at the lower (upper) end of the kinematic range.   

The correction to $2C_{2u}-C_{2d}$ from this scenario could conceivably be probed at the $\sim 3\sigma$ ($\sim 6-7\sigma$) level by SOLID (EIC). Looking past our benchmark model, such PV-DIS experiments could therefore serve as key tests for interpretation of the CDF dijet excess as resulting from a $U(1)'$ with axial couplings to quarks, though it must be noted that models with purely vectorial couplings (such as gauged baryon number) would not be probed by these measurements.

Moving beyond the scenario motivated by the CDF anomaly, we note that PV-DIS experiments possess a unique ability to probe the small $(g',M_{Z'})$ parameter space for leptophobic $Z'$ models with axial couplings to quarks. Given the relatively large shift $\Delta C_{2q}$ that may arise in this case, the observation of a significant deviation from SM expectations -- coupled with the corresponding agreement of tests of the $C_{1q}$ with the SM -- could point strongly toward a light leptophobic $Z'$ scenario. Conversely, agreement with the SM would imply severe constraints on this interesting possibility.  

\noindent{\it  Acknowledgements} 
We thank J. Erler, K. Kumar, and P. Souder for helpful discussions and Y. Li and W. Marciano for sharing their input for the $Q^2$-dependence of the $\Pi$ functions. This work was supported in part by DOE contract DE-FG02-08ER41531 (MJRM) and the Wisconsin Alumni Research Foundation (MJRM). MRB is supported by the US Department of Energy. Fermilab is operated by Fermi Research Alliance, LLC under Contract No. DE-AC02-07CH11359 with the US Department of Energy.

\end{document}